\documentclass[a4paper,12pt,english,german]{article}

\usepackage{mathtools}
\usepackage{graphicx}

\begin{document}

\centerline{\bf \Large  {The Illusory Appeal of Decoherence in the Everettian Picture:} }\smallskip
\centerline{\bf \Large {Affirming the Consequent}}
\bigskip

\centerline{ R. E. Kastner$^a$\footnote{Email: rkastner@umd.edu}}
\centerline{14 March 2016}
\smallskip

$^a$Foundations of Physics Group, University of Maryland, College
Park, USA

ABSTRACT. {\small The idea that decoherence in a unitary-only quantum theory suffices to explain
emergence of classical phenomena has been shown in the peer-reviewed literature to be seriously flawed due to circularity.  However, claims continue
to be made that this approach, also known as ``Quantum Darwinism,'' is the correct way to understand classical emergence.
This Letter reviews the basic problem and points out an additional logical problem with the argument. It is concluded that the ``Quantum Darwinism'' program fails.}

\bigskip

\large {\bf 1. Introduction.}\\

The idea that unitary-only dynamics can lead naturally to
preferred observables, such that decoherence suffices to explain
emergence of classical phenomena (e.g., Zurek 2003) has been shown in the peer-reviewed literature to be problematic.  However, claims continue
to be made that this approach, also known as 'Quantum Darwinism,' is the correct way to understand classical emergence.
 
The problem of basis ambiguity in the unitary-only theory is laid out particularly clearly by Bub, Clifton and Monton (1996), and the difficulty highlighted by them is not resolved through decoherence arguments alone. This is because decoherence is 
 relational rather than absolute (Dugi\' c and Jekni\'c-Dugi\' c 2012; Zanardi et al 2004). In order to get off the ground with a particular
structure,    ``Quantum Darwinism''-type arguments depend on
assuming special initial conditions of separable, localizable degrees of
freedom, along with suitable interaction Hamiltonians, which amount to ``seeds'' of classicality from the outset.
Under these circumstances, the purported explanation of classical emergence becomes
circular (Kastner, 2014a, 2015). But circularity is not the only problem with the decoherence-based attempt to
explain the emergence of classicality. In what follows we examine the logical structure of the argument and find a further,
serious flaw: affirming the consequent. \\

{\bf 2. The logical flaws of ``Quantum Darwinism''} \\

The structure of the Quantum Darwinism argument is as follows:\\

\noindent 
If\
1. the quantum dynamics is unitary-only, and\\
if\
2. the universe has initially separable, localizable degrees of freedom such as distinguishable atoms, and\\
if\
3. those degrees of freedom interact by Hamiltonians that do not re-entangle them, 
then\\
4. classicality emerges.\\

For decoherence to account for the emergence of
classicality under the assumption of unitary-only (U-O) evolution (approximately and only in a ``FAPP'' sense, see below), all three premises must hold.
 However, classicality is implicitly contained in 2 and 3 through the partitioning of the universal degrees of freedom into separable, localized
substructures interacting via Hamiltonians that do not re-entangle
them, so (given U-O) one has to put in classicality to get
classicality out. Premises 2 and 3 are special initial conditions
on the early universe that may not hold--certainly they are not
the most general case for an initially quantum universe. Yet it
seems common for researchers assuming U-O to assert that 2 and 3
also must hold without question. This actually amounts to the fallacy of
affirming the consequent, as follows: one observes that we have an
apparently classical world (affirm 4), and then one asserts that 1, 2
and 3 therefore must hold.

The insistence on 2 appears, for example, in Wallace's invocation
of ``additional structure on the Hilbert Space'' as ostensibly
part of the basic formalism (Wallace 2012, p. 14-15).  Such
additional structure--preferred sets of basis vectors and/or a
particular decomposition of the Hilbert space--is imposed when
quantum theory is applied to specific situations in the
laboratory. However, what we observe in the laboratory is the
already-emergent classical world, in which classical physics
describes our macroscopic measuring instruments
 and quantum physics is applied only to prepared quantum systems that are not already entangled with other (environmental) degrees of freedom.

 If the task is to explain how we got to this empirical situation from an initially quantum-only universe, then clearly we cannot
 assume what we are trying to explain; i.e., that the universe began with quasi-localized quantum systems distinguishable from each other
 and their environment, as it appears to us today. Yet Wallace includes this auxiliary condition imposing structural separability under a section entitled ``The Bare
 Formalism''
 (by which he means U-O), despite noting that we assign the relevant Hilbert space structures ``in practice'' to empirical laboratory situations. The inclusion of this sort of auxiliary condition in the ``bare formalism'' cannot be legitimate, since such imposed structures are part of the {\it application} of
the theory to {\it a particular empirical situation}. They thus constitute
contingent information, and are therefore not
 aspects of the ``bare formalism,'' any more than, for example, field boundary conditions are part of the
bare theory of electromagnetism. These separability conditions
are auxiliary hypotheses to which we cannot help ourselves,
especially since the most general state of an early quantum
universe is not one that comes with preferred basis vectors and/or
distinguishable degrees of freedom.
 Thus, the addition of this condition amounts to asserting (2), and becomes (at best) circular reasoning, or (at worst) outright affirming of the consequent, illicitly propping up the claim that quasi-classical world ``branches''  naturally appear in an Everettian (unitary-only) picture. 

 Now, to be charitable: perhaps unitary-only theorists are tacitly assuming that (1) is not subject to question; i.e. they 
 take it as a ``given.'' If one presumes the truth of (1) in this way, then (2) and (3) seem \textit{required} in order to arrive at our current apparently classical world. If (1) were really known to be true, the logical structure of
the argument would be:\  2 and 3 iff 4.\ So, rather than reject the
argument based on its circularity, such researchers seem to assume that the consequent is evidence for the truth of premises 2
and 3 (i.e., 2 and 3 together are seen as the only way that we could have arrived at the classical macro-phenomena we now experience). The possibility that the dynamics may {\it not} be wholly
unitary--the falsity of the unitary-only premise (1)--does not seem to be 
considered. However, the need to use a circular argument
in order to preserve the claims of Quantum Darwinism should prudently be taken as an indication that the U-O assumption (1) may well be false,
and that non-unitary collapse is worth exploring for a non-circular account of how classically well-defined structures arise in a world described fundamentally by quantum theory.\footnote{Such an account is proposed in Kastner (2012) and (2014b). In that account ('possibilist transactional interpretation' or PTI), decoherence can of course occur under circumstances discussed in Zurek (2003), as a deductive consequence of quantum theory under certain specified conditions; but decoherence alone is neither necessary nor sufficient as an explanation for everyday classical phenomena such as the observed determinacy of macroscopic objects. Decoherence is not necessary because classical emergence can arise through a specific collapse process in PTI, and decoherence is not sufficient because it does not solve the measurement problem (cf. Bub 1997, p. 231).} \\

{\bf 3. Conclusion.} \\

Everettian unitary-only quantum theory seems to have become  so ``mainstream''  that in many quarters it now appears to be considered the ``standard'' theory, replacing the theory consisting of Schr{\"o}dinger unitary evolution plus von Neumann non-unitary measurement transition. Yet the only way to arrive at the world of classical phenomena we experience in the unitary-only theory is to assume classicality at the outset--and even this is only approximate and ``FAPP,'' since it fails to solve the measurement problem, as noted in Bub 1997, Section 8.2. The ``decoherence'' process as invoked in service of ``Quantum Darwinism''  is at best circular and at worst amounts to the logical fallacy of affirming the consequent. The alleged utility of decoherence is greatly overstated and illusory. It is time to consider the possibility that Everett might have been wrong.

\newpage

{\bf References}

\bigskip

Bub J, Clifton R, Monton B,  1998, The Bare Theory Has No Clothes.
In {\bf Quantum Measurement: Beyond Paradox}, eds. Healey R A, Hellman
G,  {\bf Minnesota Studies in the Philosophy of Science 17}, 32-51.

Dugi\' c M., Jekni\' c-Dugi\' c J., 2012, Parallel decoherence in
composite quantum
 systems, Pramana {\bf 79}, 199

Dugi\' c M., Arsenijevi\' c M., Jekni\' c-Dugi\' c J., 2013,
Quantum correlations relativity, {\bf  Sci. China Phys., Mech. Astron.
56}, 732

Jekni\' c-Dugi\' c J.. Dugi\' c M., Francom A., 2014, Quantum
Structures of a Model-Universe: Questioning the Everett
Interpretation of Quantum Mechanics,  {\bf Int. J. Theor. Phys.
53}, 169

Kastner, R.E., 2012. {\bf The Transactional Interpretation of Quantum Mechanics: The Reality of Possibillty}.
Cambridge: Cambridge University Press.

Kastner R. E., 2014a, Einselection of pointer observables: The new
H-theorem?, {\bf  Stud. Hist. Phil. Mod. Phys. 48}, 56

Kastner R. E., 2014b, The Emergence of Spacetime: Transactions and Causal Sets,
forthcoming in {\bf Beyond Peaceful Coexistence}, I, Licata, ed.; Preprint version http://arxiv.org/pdf/1411.2072v1.pdf.

Kastner R. E., 2015, Classical selection and quantum Darwinism,
{\bf Phys. Today  68}, 8

Wallace, D., 2012, {\bf The Emergent Multiverse: Quantum Theory
according to the Everett Interpretation}. Oxford University Press,
Oxford

Zanardi P., Lidar D. A.,  Lloyd S., 2004, Quantum Tensor Product
Structures are Observable Induced, {\bf Phys. Rev. Lett.  92},
060402

Zurek W. H., 2003, Decoherence, einselection, and the quantum
origins of the classical, {\bf Rev. Mod. Phys. 73}, 715

\end{document}